\documentclass[fleqn,twoside]{article}

\usepackage[headings]{espcrc2}

\readRCS
$Id: espcrc2.tex,v 1.2 2004/02/24 11:22:11 spepping Exp $
\usepackage{graphicx}

\usepackage[figuresright]{rotating}

\newcommand{\AmS}{{\protect\the\textfont2
  A\kern-.1667em\lower.5ex\hbox{M}\kern-.125emS}}
\newcommand{\beq}{\begin{equation}}
\newcommand{\eeq}{\end{equation}}
\newcommand{\bea}{\begin{eqnarray}}
\newcommand{\eea}{\end{eqnarray}}

\title{
Non-Universal Effects in Semi-Inclusive $B$ Decays
}

\author{
U. Aglietti
\address{
Dipartimento di Fisica,\\
Universit\`a di Roma ``La Sapienza'', \\
and I.N.F.N.,
Sezione di Roma, Italy. \\[1pt]
}
\thanks{
e-mail address: Ugo.Aglietti@roma1.infn.it
}
}
       
\runtitle{ Non-Universal Effects in Semi-Inclusive $B$ Decays }
\runauthor{U. Aglietti}

\begin{document}

\begin{abstract}
We show that most spectra in the semileptonic decay $B \, \to \, X_u \, + \, l \, + \, \nu_l$, 
such as for example the distribution in the light-cone momentum $p_+ \equiv E_X - |\vec{p}_X|$
recently considered, do not have the same long-distance structure of the photon
spectrum in the radiative decay $B \, \to \, X_s \, + \, \gamma$.
On the contrary, the semileptonic distribution in the final hadron energy
$E_X$ is connected to the radiative spectrum via short-distance factors only.
The $E_X$ distribution also has a specific infrared
structure known as the ``Sudakov shoulder''.
We also discuss an explicit check of 
the resummation formula for the semileptonic decays,
based on a recent second-order computation.

\vspace{1pc}
\end{abstract}

\maketitle

\section{Introduction}


The main message of this talk is that, 
unlike what usually stated, most spectra in
charmless semileptonic $B$ decays
\beq
\label{startsl}
B \, \to \, X_u \, + \, l \, + \, \nu_l,
\eeq
such as the distribution in the light-cone momentum 
$p_+ \equiv E_X - |\vec{p}_X|$ recently considered
\cite{phatspec}, do not have the same long-distance 
structure of the photon spectrum in the radiative decays
\beq
\label{startrd}
B \, \to \, X_s \, + \, \gamma.
\eeq
We are therefore in disagreement with the conclusions in
\cite{phatspec}, in which such a short-distance relation
is instead claimed.
The reason of the non-universality is a kinematical one:
while in radiative decays (\ref{startrd}) the hard scale $Q$ 
of the hadronic sub-process is fixed to the beauty mass $m_b$, 
in the semileptonic case one integrates instead over $Q$ 
{\it up to} $m_b$. 
That modifies the structure of the infrared
logarithms in the semileptonic spectra at the 
subleading level. 

We also show that the distribution of the hadron energy $E_X$
in the semileptonic decay is connected to the rare-decay 
photon spectrum via short-distance factors only.
This semileptonic distribution also has a specific infrared
structure known as the ``Sudakov shoulder'' 
\cite{sudshould,me,noi1}.

Finally, we discuss an explicit check of the resummation formula
for the semileptonic decays, based on a recent second-order
computation, which is also a check of the basic relation between 
the hard scale $Q$ and the total hadron energy $E_X$:
\beq
\label{adnauseam}
Q \, = \, 2 \, E_X.
\eeq


The first preliminary question is whether beauty decays
can be described with perturbation theory or they do not.
Technically, the perturbative expansion is controlled by the 
QCD coupling
$\alpha\left(m_b\right) \, \cong \, 0.22 \, \ll \, 1,$
and is therefore a legitimate one.
The real problem is however: is there a quantity which shows a good
agreement with its perturbative prediction?
It is natural to consider first inclusive quantities,
which constitute ``stronger'' predictions of 
perturbative QCD (pQCD), because they are less sensitive 
to the non-perturbative hadron structure.
In other words, if perturbation theory does not work for inclusive quantities,
there is little hope that it will work for the spectra,
even after adding some non-perturbative component
--- the shape function \cite{shapefunction}.
Inclusive widths cannot be predicted with good accuracy 
because they are proportional to the fifth power of the ill-defined 
beauty mass and to $CKM$ combinations in principle unknown:
$\Gamma \propto m_b^5 | CKM |^2$.
It is therefore convenient to look at ratios of widths, 
such as the semileptonic branching fraction:
$B_{SL} \equiv \Gamma_{SL}/\Gamma_{TOT}.$
A recent second-order computation of the hadronic width
shows a good agreement with the experimental value
$B_{exp} \simeq 11\%$ \cite{new}.

Being the answer to the first question positive,
the next question concerns the expected size of 
non-perturbative effects in the spectra.
On dimensional grounds, one expects power-corrections
$\propto \Lambda/m_b \approx O(10 \%)$, where $\Lambda$ 
is a typical hadronic scale, which can be identified 
with $\Lambda_{\overline{MS}} \simeq 200$ MeV, or 
with the mass of a constituent quark $m\simeq 350$ MeV, 
or with the mass of the $\rho$ meson $m_{\rho}\simeq 770$ MeV.
We believe that a scale $\Lambda \approx 0.5$ GeV
is a reasonable choice.
Clear peaks are visible in the hadron mass distributions 
in the radiative decay and in the semileptonic one,
associated to the $K^*$ and $\rho$ final states respectively.
Since these peaks cannot intrinsically be described in pQCD,
one can take, as an estimate of the size of non-perturbative 
effects, for example the $B\to K^* \gamma$ branching fraction,
which is $\simeq 15\%$ of the total $b\to s\gamma$ rate. 

Naive physical intuition would suggest that hadronic final
states $X_s$ coming from the decay (\ref{startrd}) cannot be described 
in pQCD when the invariant mass becomes of the order of the 
hadronic scale, i.e. when $m_X \, \approx \, \Lambda$,
because of final-state hadronization effects.
This condition, as we are going to show, is actually not correct: 
non-perturbative effects enter for hadronic 
masses by far larger than the hadronic scale.
To see that explicitly, just consider the emission of a 
soft gluon by the strange quark in (\ref{startrd}). 
The invariant mass is: $m_X^2 \, = \, (p_s+k)^2 \, \simeq \, m_b \,k_+$,
where $k_+\equiv k_0+k_3$ and we have taken the strange quark flying 
in the minus direction, so that $p_s \simeq 1/2 \, m_b(1;0,0,-1)$.
The above equation shows an amplification of the soft effects
in the jet mass, because a soft momentum component
is multiplied by the hard scale.
A soft gluon with momentum of the order of the hadronic 
scale, $k_+ \approx \Lambda$, is certainly controlled by
non-perturbative effects such as the structure of the 
$B$ meson: that implies the slice
$m_X \, \approx \, \sqrt{m_b \, \Lambda}$
is non-perturbative. This is a quite distinct effect with
respect to final-state hadronization, which is expected
to be substantial in the resonance region only.
The above equation for the $m_X$ slice shows for example that in the top 
decay $t\to X_b+W$ a final jet with a mass of $\approx 10$ GeV cannot be 
described with ``pure'' pQCD!
\footnote{
For illustrative purposes, let us neglect the beauty mass and assume 
that the top width is smaller than the life-time of a hadronic
resonance.}
In general, we may identify the following kinematical regions:
\begin{enumerate}
\item
$m_X \, \approx \, Q$, i.e. the mass of the jet is of the order 
of the hard scale.
The rate can be computed in fixed-order perturbation theory
because there are no large infrared logarithms; 
\item
\label{semprepiularga}
$\sqrt{Q \, \Lambda} \, \ll \, m_X \, \ll \, Q$.
The process is still perturbative but there are large infrared
logarithms of the form
\begin{equation}
\label{infraredsi}
~~~~
\alpha^n \, \log^k \, \frac{Q^2}{m_{X}^{\,2}}
~~~~~~(k \, = \, 1,2,\cdots,2n),
\end{equation}
which need to be resummed to all orders $n$ in order
to have a consistent theoretical prediction;
\item
\label{fermiregion}
$\Lambda \, \ll \, m_X \, \le \, O\left(\sqrt{\Lambda \, Q}\right)$.
As discussed above, Fermi-motion effects,
related to soft interactions, are substantial and have to be taken into 
account by introducing a non-perturbative function, the shape function;
\item
\label{exclusiveregion}
$\Lambda \, \approx \, m_X.$
This is the exclusive or resonance region, where
the rate is dominated by few channels and
it is completely non-perturbative; to compute, one has to 
use for example lattice QCD.
\end{enumerate}
One may ask: in which sense the decay becomes 
progressively more perturbative in the limit $m_b\to\infty$?
The main peaks of the jet mass distributions, for example, occur in region 
\ref{fermiregion}. and are therefore non perturbative even in the 
infinite mass limit.
The answer to this question is that region \ref{semprepiularga}. 
becomes progressively larger in the infinite mass limit and better 
separated from the Fermi motion region \ref{fermiregion}.

\section{Non-universality}

The radiative decays (\ref{startrd}) have a simpler long-distance structure than the 
semileptonic decays (\ref{startsl}).
That is because, in the radiative case, the tree-level process is the 2-body decay
\begin{equation}
\label{radtree}
b \, \rightarrow \, s \, + \, \gamma
\end{equation}
having the large final hadron energy $2 \, E_X \, \simeq \, m_b$, 
where in the last member we have neglected the small strange mass 
$m_s \ll m_b$.
As anticipated, the total hadron energy $E_X$ fixes the hard scale $Q$ 
in the decay according to eq.~(\ref{adnauseam}), so that $Q \cong m_b$.
We are interested in the threshold region
\begin{equation}
\label{closetoborn}
m_X \, \ll \, E_X \, \le \, m_b,
\end{equation}
which can be considered a kind of ``perturbation'' of the tree-level process
due to soft-gluon effects. 
The final hadron mass $m_X$, which vanishes in lowest order, remains indeed small in higher 
orders because of the condition (\ref{closetoborn}), while the large tree-level 
hadron energy is only mildly increased by soft emissions.
 
In semileptonic decays, the tree-level process is instead the 3-body 
decay
\begin{equation}
b \, \rightarrow \, u \, + \, l \, + \, \nu
\end{equation}
and the hadron energy $E_X$ (i.e. the energy of the $up$ quark) can become
substantially smaller than half of the beauty mass $m_b/2$. 
In other words, kinematical configurations are possible with
$E_X \, \approx \, m_b/2$
as well as with $E_X \, \ll \, m_b$.\footnote{
It is clear that the condition $E_X \, \gg \, \Lambda$
must always be verified in order to deal with a hard process and 
to be justified in the use of perturbation theory.
}
Consider for example the kinematical configuration, 
in the $b$ rest frame, with the electron and the neutrino
parallel to each other, having a large hadron energy, or the configuration 
with the leptons back to back, each one with an energy $\approx \, m_b/2$,
having instead a small hadron energy.  
This fact is the basic additional complication in going from the radiative 
decays to the semileptonic ones: the hard scale is no more fixed by the
heavy flavor mass but it depends on the kinematics according to
eq.~(\ref{adnauseam}).

Before going on, let us give the general definition of a short-distance
quantity in pQCD as far as threshold effects are concerned.
We consider a process characterized by a hard scale 
$Q \, \gg \, \Lambda,$
and by the infrared scales 
$m_X^4/Q^2, ~ m_X^2  \,\, \ll \,\, Q^2$  
--- the soft scale and the collinear scale respectively.
If infrared effects are absent in the quantity under consideration, 
logarithmic terms of the form (\ref{infraredsi}) must not appear 
in its perturbative expansion.
The presence of  such terms would indeed signal 
significant contributions from small momentum scales.
That is because these terms originate from the integration of the infrared-enhanced 
pieces of the QCD matrix elements from the hard scale down to one of the 
infrared scales:
\beq
\label{integrali}
\int_{m_X^4/Q^2}^{Q^2} \, \frac{dk_{\perp}^2}{k_{\perp}^2},
~~~
\int_{m_X^2}^{Q^2} \, \frac{dk_{\perp}^2}{k_{\perp}^2}
~~~ \Rightarrow ~~~
\log \, \frac{Q^2}{m_X^2}.
\eeq
If perturbation theory shows a significant contribution from small
momentum scales to some cross section or decay width, we believe there
is no reason to think that the same should not occur in a non-perturbative 
computation.

On the contrary, a quantity such as a spectrum or a ratio of spectra
is long-distance dominated if it contains infrared logarithmic terms in the perturbative 
expansion of the form (\ref{infraredsi}).
These terms must be resummed to all orders of perturbation theory in the threshold region 
(\ref{closetoborn}).
Our criterion to establish whether a quantity is short-distance or it is not
is rather ``narrow'': we believe it is a fundamental one and we use it 
systematically, i.e. we derive all its consequences. 
The consequences, as we are going to show, are in some cases not trivial.

Factorization and resummation of threshold logarithms in the radiative 
decays (\ref{startrd}) leads to an expression for the event fraction
of the form:
\bea
\label{singola}
\frac{1}{\Gamma_r} \int_0^{t_s} \frac{d\Gamma_r}{d t_s'} dt_s' 
&=& C_r\left[\alpha(m_b)\right] \, \Sigma\left[t_s; \alpha(m_b)\right] 
\nonumber\\
&& ~~~ + \, D_r\left[t_s;\alpha(m_b)\right],
\eea
where $\Gamma_r$ is the inclusive radiative width and 
$t_s \equiv m_{X_s}^2/m_b^2$ 
($x_{\gamma} \equiv 2E_{\gamma}/m_b = 1-t_s$).
We have defined the following quantities, all having an expansion in powers 
of $\alpha$:
\begin{itemize}
\item
$C_r\left( \alpha \right)$, a short-distance, process-dependent 
coefficient function, independent on the hadron variable $t_s$.
The explicit expression of the first-order correction 
$C_{\, r}^{(1)}$ has been given in \cite{me,acg};
\item
$\Sigma(u;\,\alpha)$, the universal QCD form factor for heavy 
flavor decays, having a double expansion of the form:
\bea
\label{Sigma}
& & \Sigma[u;\,\alpha] \, = \, 
1 \, + \, \sum_{n=1}^{\infty} \sum_{k=1}^{2n} \, 
\Sigma_{nk} \, \alpha^n \, \log^k \frac{1}{u}
\nonumber
\\
& = & 1 
- \frac{1}{2} \frac{\alpha\,C_F}{\pi} \log^2 \frac{1}{u}
+ \frac{7}{4} \frac{\alpha\,C_F}{\pi} \log \frac{1}{u} +
\nonumber\\
&+& \frac{1}{8} \left(\frac{\alpha\,C_F}{\pi}\right)^2 \log^4 \frac{1}{u}
+ \cdots,
\eea
where $C_F = 4/3$.
In higher orders, as it is well known, $\Sigma$ contains at most 
two logarithms for each power of $\alpha$, coming from the overlap 
of the soft region and the collinear one in each emission
and it has an exponential structure;
\footnote{
This form factor can be expressed in terms of the shape function $f$
via a universal coefficient function $\tilde{C}$ 
(not to be confused with $C_r$):
$\Sigma[u; \, \alpha(Q)] \, = \, \tilde{C}(u; \, Q,\mu_F)*f(u; \, \mu_F)$,
where the star denotes a convolution and $\mu_F \le Q$ is a factorization
scale. The coefficient function $\tilde{C}$ described hard collinear 
emissions and soft emissions with energies between the factorization
scale $\mu_F$ and the hard scale $Q$.
If we take for example $\mu_F=Q$, the first integral 
and the second one on the l.h.s. 
of eq.~(\ref{integrali}) are related to the shape function and to its 
coefficient function respectively \cite{shapefunctionren}.
}
\item
$D_r\left(t_s; \, \alpha \right)$, a short-distance,
process-dependent remainder function, not containing infrared logarithms 
and vanishing for $t_s \, \rightarrow \, 0$ as well as for 
$\alpha \, \rightarrow \, 0$.
\end{itemize}

\noindent
Factorization and resummation of threshold logarithms in the semileptonic 
case is conveniently made starting with distributions not integrated over
$E_X$, i.e. not integrated over the hard scale $Q$.
The most general distribution in process (\ref{startsl}),
a triple distribution, has been originally resummed in 
\cite{me} (see also \cite{noi1,noi2,noi3} and \cite{gardifinal}):
\bea
\label{tripla}
&& \frac{1}{\Gamma} \int_0^u \frac{d^3\Gamma}{dx dw du'} du' 
= C\left[x,w;\alpha(w m_b)\right] ~ {\rm x}
\nonumber\\
&& ~ {\rm x} ~ \Sigma\left[u;\alpha(w m_b)\right]
+ D\left[x,u,w;\alpha(w\,m_b)\right],
\eea
where we have defined the following kinematical variables:
\bea
x \, = \, \frac{2 E_l}{m_b},~~
w \, = \, \frac{Q}{m_b}~~(0\le w \le 2)
\eea
and
\bea
u = \frac{1 - \sqrt{1 - \left(2m_X/Q\right)^2} }{1 + \sqrt{1 - \left(2m_X/Q\right)^2} }
\, \simeq \, \left(\frac{m_X}{Q}\right)^2.
\eea
In the last member we have kept the leading term in the threshold region $m_X \ll Q$ only
and $\Gamma$ is the total semileptonic width.
Eq.~(\ref{tripla}) is a ``kinematical'' generalization of eq.~(\ref{singola}) 
together with the quantities involved:
\begin{itemize}
\item
$C\left[x,w;\alpha\right]$, a coefficient function, dependent on the 
hadron energy and lepton one but independent on the hadron variable $u$;
\item
$\Sigma[u;\,\alpha(w \, m_b)]$, the universal QCD form factor for heavy 
flavor decays, which now is evaluated for a coupling with a general
argument $Q \, = \, w \, m_b \le m_b$;
\item
$D\left[x,u,w;\alpha\right]$, a remainder function, depending on all 
the kinematical variables, not containing infrared logarithms and 
vanishing for $\alpha \to 0$ and $u\rightarrow 0$.
\end{itemize}
Many semileptonic spectra, such as the $p_+$ spectrum,
are obtained integrating the resummed triple differential distribution 
(\ref{tripla}) over the hadron energy, i.e. over the hard scale of the 
hadronic subprocess.
The infrared logarithms entering the distribution in
$\hat{p}_+ \equiv p_+/m_b$ can be resummed by means of an effective 
form factor of the form \cite{noi3}
\bea
\Sigma_P
\left[
\hat{p}_+; \, \alpha(m_b)
\right]
&\propto& 
\int_0^1 dw \, C_H\left[w; \, \alpha(w \, m_b)\right] \, {\rm x}
\nonumber\\
& & {\rm x} ~
\Sigma \left[
\hat{p}_+/w; \, \alpha(w \, m_b)
\right],
\eea
where we have omitted a constant factor and $C_H(w;\alpha) = 2w^2(3-2w)+O(\alpha)$ 
is a coefficient function.
There is an integration over $w\in[0,1]$, which enters
the first argument of the universal form factor $\Sigma$ 
as well as the argument of the QCD coupling.
Hadronic subprocesses with different hard scales $Q = w \, m_b$
therefore contribute. 
By measuring the photon spectrum in the radiative decay 
(\ref{startrd}), one can only extract $\Sigma[u;\alpha(m_b)]$,
while for the $\hat{p}_+$ spectrum it is necessary to know 
$\Sigma[u; \, \alpha(w \, m_b)]$ with $0 \le w \le 1$.
We conclude therefore that it is not possible to derive the 
long-distance structure of the $\hat{p}_+$ spectrum from 
the radiative decay (\ref{startrd}). 
Furthermore, according to the criterion discussed before,
the ratio between the $\hat{p}_+$ spectrum and the radiative mass
distribution evaluated for $t_s = \hat{p}_+$ is not a 
short-distance quantity: infrared logarithms only cancel in leading
logarithmic approximation.
The relation between the $\hat{p}_+$ spectrum and the radiative one 
therefore is not a true short-distance relation.
These non-universality effects are of higher order in the log 
counting but not in the twist. 
Our conclusions are therefore in disagreement with those ones
derived in \cite{phatspec}.
Similar considerations can be repeated for the charged lepton
spectrum and for the semileptonic distributions in the hadronic mass 
normalized to the hadronic energy or to the beauty mass \cite{noi2}.

\section{Hadron energy spectrum}

The hadron energy spectrum contains in first order
logarithmic terms for $w>1$ 
of the form \cite{enspec,ndf}
\beq
\label{sudshoulder}
\alpha \log^2 (w-1)~~~{\rm and}~~~\alpha \log (w-1),
\eeq
which formally diverge for $w\to 1^+$, inside the
physical domain.
These terms originate from the fact that at tree level
the hadron energy is restricted to half the beauty mass:
\beq
\label{limitation}
E_X^{(0)} = E_u \le \frac{m_b}{2},
\eeq
while it that go up to the whole beauty mass in higher
orders:
\beq
E_X \le m_b.
\eeq
Slightly above $m_b/2$ there are soft-gluon effects
related to real emissions which cannot be cancelled
by the virtual corrections, subject to the limitation
(\ref{limitation}): that is the physical origin of 
the large logarithms in (\ref{sudshoulder}).

\noindent
For $w\le 1$ there are no large logarithms and therefore
the usual fixed-order expansion holds:
\beq
\label{smaller}
\frac{1}{\Gamma}\frac{d\Gamma}{dw}
= L^{(0)}(w) + \alpha L^{(1)}(w) + \cdots
\eeq
For $w>1$ the large logarithms can be resummed by means
of an expression of the form \cite{noi1}:
\bea
\label{larger}
\frac{1}{\Gamma}\frac{d\Gamma}{dw} &=&
C_{W1}(\alpha)
\Big\{
1 - C_{W2}(\alpha)\Sigma\left[w-1;\alpha(m_b)\right]
\nonumber\\
&& ~~~~~~ + \, D_W(w;\alpha)
\Big\}.
\eea
The following quantities enter the resummed spectrum:
\footnote{The explicit expressions for the coefficient
functions and the remainder function can be
found in \cite{noi1}.}
\begin{enumerate}
\item
$C_{W1}(\alpha)$, a first coefficient function
determined imposing the continuity of the spectra 
(\ref{smaller}) and (\ref{larger}) for $w\to \mp 1$
respectively;
\item
$C_{W2}(\alpha)$, a second coefficient function,
determined imposing the consistency of the resummed
expression with the fixed-order computation;
\item
$\Sigma\left[w-1;\alpha(m_b)\right]$, the 
universal QCD form factor entering the hadron mass 
distribution in the radiative case (\ref{startrd}),
resumming all the large logarithmic corrections 
for $w \to 1^+$, such as those in (\ref{sudshoulder});
\item
$D_W(w;\alpha)$, a remainder function, starting
to $O(\alpha)$, free from infrared logarithms 
and vanishing for $w\to 1^+$.
\end{enumerate}
By measuring the hadron energy spectrum slightly
above $m_b/2$ one can determine the universal
QCD form factor $\Sigma[w-1;\alpha(m_b)]$,
which also enters the radiative spectrum.
Unlike previous cases, the relation between 
the semileptonic hadron energy spectrum and the 
radiative spectrum is a true short-distance 
relation.

A form analogous to (\ref{larger}) 
is also found in the resummation of the $C$-parameter
above the 3-jet threshold \cite{private}.
The resummation of the double distribution
in the hadron energy and in the charged lepton one 
involves a generalization of (\ref{larger}) 
\cite{noi1}.

\section{Check of resummation formula}

The resummed formula on the r.h.s. of eq.~(\ref{tripla}) 
has been derived with general effective-theory arguments, 
holding to any order in $\alpha$ \cite{me,noi1}: 
one basically considers the infinite-mass limit for the beauty quark,
$m_b \, \to \, \infty,$ while keeping the hadronic energy $E_X$ and 
the hadronic mass $m_X$ fixed.
The main result is that the hard scale $Q$ enters:
\begin{enumerate}
\item
\label{prop1}
the argument of the infrared logarithms factorized in the QCD form
factor $\Sigma[u;\,\alpha]$, 
$\log 1/u \, \cong \, \log  Q^2/m_X^2;$ 
\item
\label{prop2}
the argument of the running coupling $\alpha=\alpha(Q)$, from which 
the form factor $\Sigma[u;\,\alpha(Q)]$ depends --- as well as the 
coefficient function and the remainder function do.
\end{enumerate}
An explicit check of property \ref{prop1}. has been obtained
verifying the consistency between the resummed expression 
on the r.h.s. of eq.~(\ref{tripla}) expanded up to $O(\alpha)$ 
and the triple distribution computed to the same order in \cite{ndf}.
Since the dependence of the coupling on the scale is a 
second-order effect, point \ref {prop2}. cannot be verified
with the above computation.
The possibility for instance that the hard scale $Q = w\, m_b$ 
in (\ref{tripla}) is fixed instead by the beauty mass $m_b$,
i.e. by $Q \, = \, m_b$,
cannot be explicitly ruled out by comparing with the $O(\alpha)$ triple
distribution, because
$\alpha(w \, m_b) \, = \, \alpha(m_b) \, + \, O\left(\alpha^2\right).$
As far as we known, the only available second-order computation is that of the
$O(\alpha^2 n_f)$ corrections to the distribution in the light-cone
momentum $\hat{p}_+$ \cite{phatspec}.
Expanding our resummed expression for the $\hat{p}_+$ spectrum up to second 
order and comparing the $O(n_f)$ parts of the logarithmic coefficients, we have
found complete agreement with the explicit computation. 
If the coupling is evaluated instead for example at the beauty mass, 
we obtain different values of the coefficients, in disagreement
with the Feynman diagram evaluation
\footnote{
A few days after paper \cite{noi3} was put on spires, 
\cite{gardifinal} also appeared, whose results on the $\hat{p}_+$ 
logarithmic coefficients are in agreement with \cite{noi3}.}.

\section{Conclusions}
We have shown that there is no universality of
long-distance effects between the radiative decay
photon spectrum and many semileptonic decay 
distributions, such as for example the distribution
in the light-cone momentum $p_+$.
An exception is represented by the semileptonic distribution
in the hadron energy $E_X$, which has exactly the same infrared
structure of the radiative spectrum.
These conclusions are reached by means of an all-order 
resummation of the threshold logarithms.

\end{document}